\documentclass[
    aps,prl,twocolumn,
    reprint,
    superscriptaddress,
    floatfix,
    amssymb,
    longbibliography
]{revtex4-1}

\usepackage{amsmath}
\usepackage{amsfonts}
\usepackage{graphicx}

\usepackage[utf8]{inputenc}

\usepackage[normalem]{ulem}

\usepackage{xcolor}

\definecolor{linkColor}{rgb}{0,0.3,0.7}
\usepackage{hyperref}
\hypersetup{colorlinks=true,
            allcolors=linkColor,
            pdfborder={0 0 0},
            pdfencoding = auto
            }

\begin{document}

\title{Subdiffusive Activity Spreading in the Diffusive Epidemic Process}
\author{Borislav Polovnikov}
\author{Patrick Wilke}
\author{Erwin Frey}
\email{frey@lmu.de}

\affiliation{
Arnold Sommerfeld Center for Theoretical Physics and Center for NanoScience, Department of Physics, Ludwig-Maximilians-Universit\"at M\"unchen, Theresienstrasse 37, D-80333 Munich, Germany
}

\begin{abstract}
The diffusive epidemic process is a paradigmatic example of an absorbing state phase transition in which healthy and infected individuals spread with different diffusion constants.
Using stochastic activity spreading simulations in combination with finite-size scaling analyses we reveal two qualitatively different processes that characterize the critical dynamics: subdiffusive propagation of infection clusters and diffusive fluctuations in the healthy population. This suggests the presence of a strong-coupling regime and sheds new light on a longstanding debate about the theoretical classification of the system.
\end{abstract}

%\date{\today}

\maketitle

Absorbing state phase transitions are an important class of collective phenomena in nonequilibrium physics, and their analysis has led to many conceptual advances that are also important for more detailed and realistic models~\cite{Hinrichsen.Hinrichsen.2000, henkel2008non, Odor.Odor.2004, Taeuber.Vollmayr-Lee.2005}.
Here we study the \textit{Diffusive Epidemic Process} (DEP), a stochastic many-body system introduced to conceptually model the propagation of an epidemic in a fluctuating population~\cite{kree1989, woh1998}.
It is defined as a reaction-diffusion model on a lattice with two different particle types $A$ and $B$ which diffuse independently with diffusion constant $D_A$ and $D_B$.
The reactions, $A + B \rightarrow 2 \, B$ and $B \rightarrow A$, can be interpreted as an infection of a `healthy' individual ($A$) by a `sick' one ($B$) with infection rate $\lambda$, and recovery of sick individuals with a typical recovery time $\tau$. 
Interestingly, it can also be viewed as a minimal model for cell polarity~\cite{Altshuler_etal:2008, Brauns.Frey.2020}.
Compared to the epidemic process described by the prominent directed percolation~\cite{Hinrichsen.Hinrichsen.2000}, there are two important conceptual differences, namely that it contains two different types of particles instead of only one, and that the reactions preserve the total number of particles. 
Both, as we will show, have important consequences for the critical dynamics at the phase transition from an active state with a finite fraction of $B$-particles to an absorbing state in which $B$-particles are absent. 

While we are interested in the full stochastic dynamics of the DEP, it is instructive to first consider the corresponding mean-field description in terms of (mass-conserving) reaction-diffusion equations~\cite{Frey.Brauns.2020}
\begin{subequations}
\label{eq:mean_field}
\begin{align}
    \partial_t \, a (x,t) &=
    D_A \nabla^2 a - \lambda \, a \, b + b / \tau
    \, , \\
    \partial_t \, b (x,t) &=
    D_B \nabla^2 b + \lambda \, a \, b - b / \tau 
    \, .
\end{align}
\end{subequations}
Here $a (x,t)$ and $b (x,t)$ denote the time-dependent local densities of individuals of type $A$ and $B$, respectively. 
The total particle density $n(x,t) \,{=}\, a(x,t) \,{+}\, b(x,t)$ plays a special role as its spatial average $\rho$ is a conserved quantity and hence can serve as a control parameter for the system's behaviour.  
Upon decreasing the average total density $\rho$, this mean-field theory shows a transition (transcritical bifurcation) from an active state with a finite (average) density of sick individuals to an absorbing state with $b (x,t) \,{=}\, 0$ at $\rho_c^\text{mf} \,{=}\, (\lambda \tau)^{-1}$.
%For high total density $\rho$, the disease persists forever, whereas for low density all individuals eventually recover and the system remains in a disease-free absorbing state. 

%Previous analytical and numerical results.
The full stochastic model is formulated in terms of a master equation that can be mapped to a field theory for the corresponding particle densities; for reviews see e.g.\/ Refs.~\cite{Taeuber.Vollmayr-Lee.2005, wiese2016coherent, weber2017master}. 
This field theory, which is a genuine extension of the Reggeon field theory describing directed percolation \cite{Cardy1980}, serves as a starting point for a renormalisation group (RG) analysis of the dynamics in the vicinity of the absorbing state phase transition~\cite{Taeuber.Vollmayr-Lee.2005}. 
In such RG studies, three different universality classes have been identified, depending on the relative size of the diffusion constants~\cite{kree1989, woh1998}.
While the cases $D_A \,{=}\, D_B$ and $D_A \,{<}\, D_B$ were both amenable to a perturbative RG calculation close to the upper critical dimension $d_\text{uc} \,{=}\, 4$, the (biologically more relevant~\cite{Altshuler_etal:2008, Brauns.Frey.2020}) case $D_A \,{>}\, D_B$ proved to be theoretically challenging and puzzling, since neither perturbative~\cite{woh1998} nor non-perturbative methods~\cite{wschebor2017nprg} have yet found a stable fixed-point structure.
The absence of a RG fixed point was initially interpreted as evidence for a fluctuation-induced discontinuous phase transition~\cite{oerding2000fluctuation}.
However, later numerical simulations clearly indicated a continuous transition, but largely disagreed on the values for the critical exponents~\cite{Freitas.Hilhorst.2000, fulco2001montePhysRevE, bertrand2007critical, maia2007diffusive, maia2008diffusive, corso2010critical, argolo2019stationary}. 

%Key results.
Hence, the nature of the absorbing state phase transition, the values of the critical exponents, and their relation to the limiting case of equal diffusion constants remain important open questions, which we address here by means of large-scale numerical simulations of the one-dimensional DEP employing a Gillespie algorithm~\cite{Gillespie2007,GibsonBruck2000_NextReactionMethod}.
%Simulation approach and observables.
There are two complementary ways to simulate the critical dynamics of absorbing state phase transitions~\cite{hinrichsen2000non}. 
In simulations studying the dynamic spreading of activity~\cite{grassberger_torre_1979, Hinrichsen.Hinrichsen.2000, lubeck2004universal, henkel2008non} 
one initializes seeds of `sick' individuals ($B$-particles) and then statistically analyses the dynamics of the ensuing $B$-clusters. 
This approach has previously been employed to obtain highly accurate values for a variety of systems including directed percolation~\cite{grassberger_1989_DP_2plus1, jensen_fogedby_1990, jensen_1991_univ_class_1dim, jensen_1992_criticality_threedim_CP, grassberger_1996_self_organized,voigt_ziff_1997}, pair contact processes~\cite{noh_park_2004}, branching and annihilating random walks~\cite{jensen_1994_crit_exp_BARW}, and the triplet annihilation model~\cite{dickman_1990_noneq_criticality_triplet}.
The most important observables are the mean number of $B$-particles $\langle N_B (\rho,t)\rangle$ in the cluster, the survival probability $P_\text{surv}(\rho,t)$ of these clusters and their mean-square radius $\langle R^2(\rho,t) \rangle$, conditional on survival, as a function of total particle density $\rho$ and time $t$.
At a continuous absorbing state phase transition the following scaling relations hold~\cite{lubeck2004universal,henkel2008non}
\begin{subequations}
\label{eq:initial_seed_observables}
\begin{align}
    \langle N_B (\rho,t) \rangle 
    &= t^{\theta} \, \widehat N_B 
       \bigl( 
       (\rho{-}\rho_c)^{\nu_\parallel} \, t
       \bigr) 
       \, , \\
    P_\text{surv}(\rho,t) 
    &= t^{-\delta} \, \widehat P_\text{surv}
       \bigl( 
       (\rho{-}\rho_c)^{\nu_\parallel} \, t 
       \bigr) \, , \\
    \langle R^2(\rho,t) \rangle 
    &= t^{2/z} \, \widehat {R^2}  
      \bigl( 
      (\rho{-}\rho_c)^{\nu_\parallel} \, t 
      \bigr) \, ,  
\end{align}
\end{subequations}
where $\rho_c$ is the critical total density, and $\theta$, $\nu_\parallel$, $\delta \,{=}\, \beta'/\nu_\parallel$, and $z \,{=}\, \nu_\parallel/\nu_\perp$ are independent critical exponents. 
Alternatively, one may simulate the dynamics starting from a system homogeneously filled with $B$-particles. 
In this case one expects a similar scaling relation for the survival probability as above, whereas 
the average density of sick individuals, $\left< \rho_B^{} (\rho,t) \right>$, then obeys the scaling form~\cite{Hinrichsen.Hinrichsen.2000, henkel2008non}
\begin{equation}
\label{eq:homogeneous_initial_state}
    \langle \rho_B^{} (\rho,t)\rangle 
    = t^{-{\beta}/{\nu_\parallel}} \, 
      \widehat \rho_B^{}
      \bigl(  (\rho{-}\rho_c)^{\nu_\parallel} \, t 
      \bigr) \, ,
\end{equation} 
with an exponent $\beta$ that in general is different from $\beta'$. We will measure time in units of the recovery time, setting $\tau \,{=}\, 1$, and length in units of the lattice spacing. For specificity, we fix the infection rate to $\lambda \,{=}\, 0.2$; see the SM~\cite{Note:SM} for details of the numerical implementation.

%Equal diffusion constants.
We start our analysis with the case of equal diffusion constants, $D \,{=}\, D_{A,B}$. Then, the dynamics greatly simplifies as the total density $n (x,t) \,{=}\, a(x,t) \,{+}\, b(x,t) $ strictly obeys a diffusion equation, $\partial_t n \,{=}\,  D \nabla^2 n$, and there are other general implications for the critical dynamics: The absence of propagator renormalisations in  perturbative RG analyses~\cite{kree1989, woh1998} suggests that the dynamic exponent equals $z \,{=}\, 2$, i.e.\ the dynamics is \textit{diffusive}, and rapidity reversal symmetry implies $\beta \,{=}\, \beta'$ \cite{henkel2008non}.  

\begin{figure}[!b]
\centering
\includegraphics[width=\columnwidth]{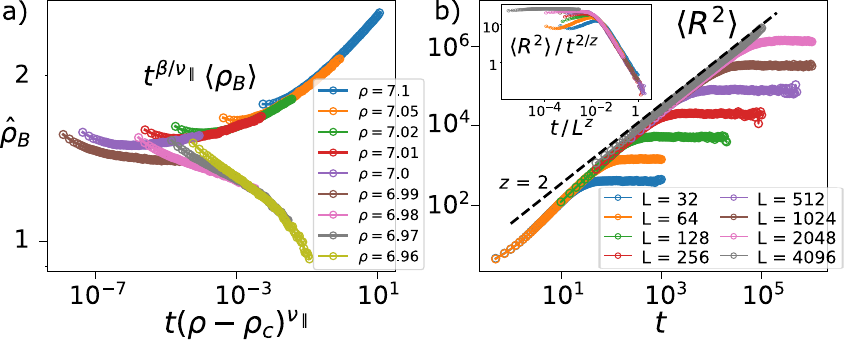}
\caption{
\textbf{Scaling analysis for equal diffusion constants.} 
(a) Scaling plot for the density of $B$-particles $\langle \rho_b \rangle$ using Eq.~\ref{eq:homogeneous_initial_state}. 
(b) Mean-square cluster radius $\langle R^2 \rangle$ at the critical density $\rho_c$ and the corresponding scaling plot with $z \,{=}\, 2$ (inset); a typical simulation run is shown in Movie1.  The diffusion constant is set to $D\,{=}\,1$.
}
\label{fig:equal_diffusion}
\end{figure}

Figure~\ref{fig:equal_diffusion}a shows our simulation results for the total density $\left< \rho_B \right>$ of $B$-particles, starting from a spatially uniform state.
A scaling analysis of the raw data for $\left< \rho_B \right>$  using Eq.~\eqref{eq:homogeneous_initial_state}  yields the critical density $\rho_c \,{=}\, 6.995(10)$, and the critical exponents $\beta/\nu_\parallel \,{=}\, 0.087(6)$ and $\nu_\parallel \,{=}\, 4.05(40)$, with the error margins estimated from the breakdown of a reasonable data collapse; see SM~\cite{Note:SM}. 
At the critical density $\rho_c$, the mean-square radius $\langle R^2 \rangle$ of a cluster of $B$-particles seeded at the origin exhibits power-law behaviour $\langle R^2 \rangle \,{\sim}\, t^z$ with $z \,{=}\, 2.00(4)$ which saturates at a characteristic time scale $t_b(L)$ that depends on the system size $L$ (Fig.~\ref{fig:equal_diffusion}b).
Finite size scaling is consistent with $\langle R^2 \rangle (t,L) \,{=}\, t^{2/z} F (t/L^z) 
$, i.e.\ the saturation of cluster growth at $t_b(L) \,{\sim}\, L^z$ shows the same scaling behaviour as cluster growth itself.
This is precisely what one would expect if the critical dynamics of the DEP is characterised by a single time scale, which conclusively shows that the epidemic process for equal diffusion constants is diffusive, consistent with earlier simulation results \cite{maia2007diffusive,corso2010critical}  and RG analyses~\cite{kree1989,  woh1998, janssen2001comment}.   
Within the error margins, the above results are also consistent with the correlation length exponent $\nu_\perp \,{=}\, 2 / d$, which was argued to be exact due to symmetries of the underlying field theory~\cite{woh1998,janssen2001comment}, but has recently been questioned by a non-perturbative functional RG study~\cite{wschebor2017nprg}. 

Finally, we performed a scaling analysis of $\langle N_B \rangle$ and $P_\text{surv}$ and find $\theta \,{=}\, 0.32(2)$ and $\beta' \,{=}\, \beta$, respectively; see SM~\cite{Note:SM}. These results affirm the rapidity reversal  symmetry and the ensuing hyperscaling relation $\theta \nu_\parallel \,{=}\, d \nu_\parallel / z - \beta - \beta'$; see Refs.~\cite{Hinrichsen.Hinrichsen.2000}.

\begin{figure}[!t]
\centering
\includegraphics[width=\columnwidth]{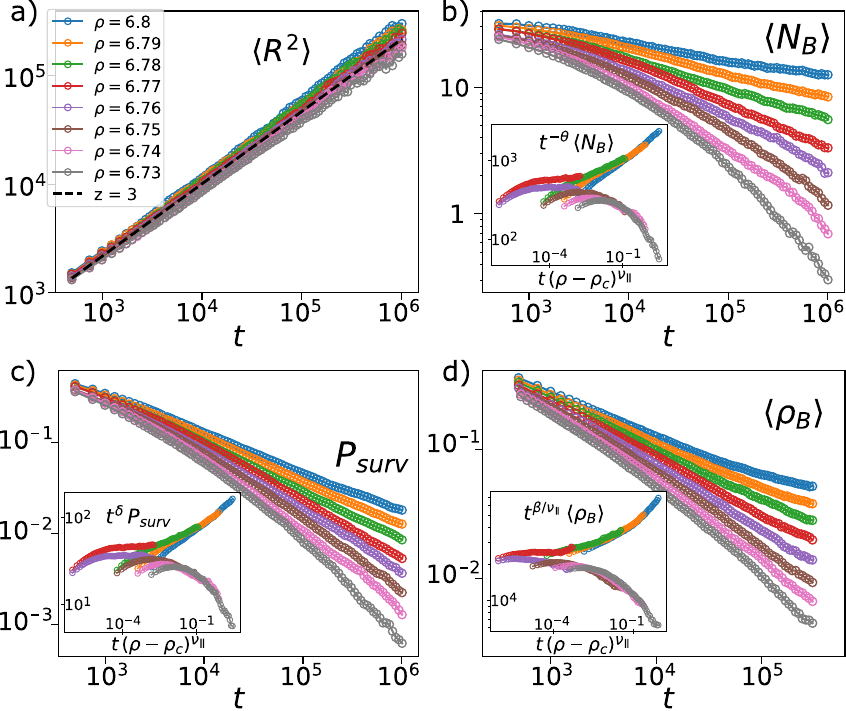}
\caption{ 
\textbf{Scaling analysis for unequal diffusion constants, $D_A > D_B$.}
Simulation data for a set of total densities $\rho$ as indicated in the graph (different colors): 
(a-c) The mean-square radius $\langle R^2 \rangle$, mean number $\langle N_B \rangle $ of $B$-particles, and survival probability $P_\text{surv}$ of a cluster evolving from an initial seed of $B$-particles placed in the center of the lattice, respectively. 
The straight dashed line in (a) indicates a power-law growth $\langle R^2 \rangle \,{\sim}\, t^{2/z}$ with the dynamical exponent $z \,{=}\, 3.0$. 
The insets show scaling collapses using Eq.~\ref{eq:homogeneous_initial_state}.
(d) The mean $B$-particle density in a system starting from a homogeneously filled lattice with the inset showing  
the scaling collapse using Eq.~\ref{eq:homogeneous_initial_state}.
The system size is $L = 4\,096$ and the ensemble size is $50 \, 000$ and $4\,000$ for simulations starting from a seed (a-c) or a spatially uniform system (d) respectively. 
For typical simulation runs starting from a seed or homogeneous state please see Movie2 and Movie3.
%\bnote{es gab einzelne runs z.B. fuer rho=6.77 , 6.78 und 6.79 mit L=8192, alle anderen hatten L=4096. rho=6.80 hatte ein ensemble mit 30000 statt 50000, da es schon weit in der aktiven Phase ist und weniger cluster aussterben}
}
\label{fig:scaling_unequal_diff}
\end{figure}

%Unequal diffusion constants.
Next, we studied the case where the sick individuals diffuse more slowly than the healthy ones, $D_B \,{<}\, D_A$. 
For the time being, we set the diffusion constants to $D_A \,{=}\, 1$ and $D_B \,{=}\, 0.5$, which corresponds to a situation where, during a typical recovery period, the mean-square distance traveled by both types of particles is of the order of the lattice spacing. 

All of our data (Fig.~\ref{fig:scaling_unequal_diff}) are consistent with the scaling forms given by Eqs.~\eqref{eq:initial_seed_observables} and \eqref{eq:homogeneous_initial_state}, clearly showing that the corresponding absorbing state phase transition is continuous.
In the activity spreading simulations, we get the best statistics for the mean number $\langle N_B \rangle$ of $B$-particles in the cluster and the cluster survival probability $P_\text{surv}$, since these quantities are obtained by averaging over all $50 \, 000$ realisations (Fig.~\ref{fig:scaling_unequal_diff}b,c). 
From a scaling analysis of these simulation data we determine accurate estimates of the critical density $\rho_c  \,{=}\,  6.765(5)$
%\footnote{The mean-field critical density is $\rho_c^\text{mf} \,{=}\, 5$, i.e.\ fluctuations renormalise that value such that one needs a higher density to reach the active state.}
and the critical exponents $\delta \,{=}\, {\beta'}/{\nu_\parallel} \,{=}\, 0.66(3)$, $\theta \,{=}\, {-}0.38(4)$, and  $\nu_\parallel \,{=}\, 3.8(5)$. 
The scaling analysis of the mean density of $B$-particles $\langle \rho_B \rangle$, obtained from simulations starting from a homogeneous initial condition, confirms the estimate for the critical density and yields the estimates ${\beta}/{\nu_\parallel} \,{=}\, 0.47(3)$ and $\nu_\parallel \,{=}\, 3.8(5)$ (Fig.~\ref{fig:scaling_unequal_diff}d).
The value for the exponent ${\beta}/{\nu_\parallel}$ is in accordance with earlier computational studies analysing the steady states ~\cite{maia2007diffusive}, while the exponents $\delta$ and $\theta$ have not been measured previously.
%\enote{Is it correct to say that these numerical studies analyse the steady states and do we cite all the pertinent references here?} %\bnote{This is the only study which measured $\rho_B$ vs t as a double check for their value of $\rho_c$, apart from that they used a restricted steady state (i.e. systems are not allowed to die) approach}. 
In particular, our data for the survival probability in the dynamic spreading simulations clearly show that for $D_A \,{\neq}\, D_B$ the rapidity reversal symmetry is broken ($\beta \,{\neq}\, \beta'$) as expected from field theory \cite{woh1998}.

Surprisingly, our activity spreading simulations show that the mean-square cluster radius at the critical density $\rho_c$ grows \textit{subdiffusively} $\langle R^2 \rangle \,{ \sim}\, t^{2/z_s}$ with a dynamic exponent $z_s \,{=}\, 3.0(1)$. 
This appears to be in conflict with previous simulations that obtained $z \,{\approx}\,  2$ by measuring the average extinction time of homogeneously initialized lattices~\cite{maia2007diffusive, corso2010critical}.
Perturbative RG studies also find $z \,{=}\, 2$~\cite{woh1998,janssen2001comment}: it is argued that in a perturbative calculation, independent of $D_A$ and $D_B$, there are no terms that would renormalise the diffusion terms.
This indicates that our numerical results point to a strong coupling behaviour that is inaccessible to perturbation theory.
\begin{figure}[t]
\centering
\includegraphics[width=\columnwidth]{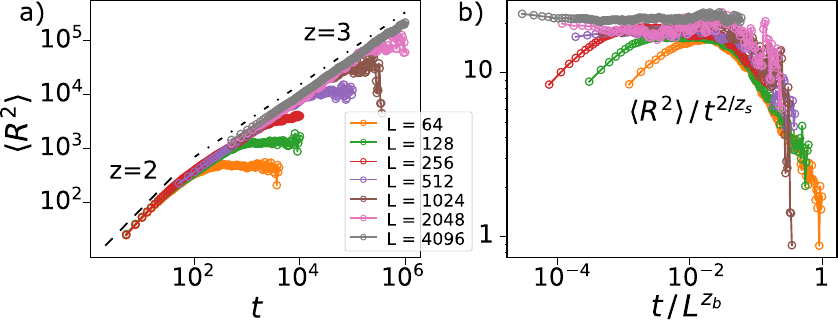}
\caption{
\textbf{Finite sizes scaling for cluster spreading at criticality for $D_A > D_B$.} Dynamic spreading simulation data for $\langle R^2 \rangle (t)$ at the critical density $\rho_c \,{=}\, 6.765 $ for varying system sizes $L$ indicated in the graph. (a)  The raw data for the mean-square radius shows a crossover from diffusive growth ($z_s \,{=}\, 2$) to asymptotic subdiffusive spreading ($z_s \,{\approx}\, 3$)  before it saturates at $\langle R^2 \rangle \sim L^\alpha$ with $\alpha \,{=}\, 2 z_b/z_s \,{\approx}\, 4/3$. 
(b) Finite size scaling analysis using the scaling law given in Eq.~\eqref{eq:finite_size_scaling_R2}. The scaling collapse is best for $z_b \,{=}\, 2.0(1)$ and $z_s \,{=}\, 3.0(1)$.
The parameters of the simulations were as specified in the main text with an ensemble size of $50 \, 000$. 
}
\label{fig:mean-square_radius_fss}
\end{figure}
To investigate this further, we now resort to a finite-size scaling analysis for the mean-square cluster radius at the critical density  (Fig.~\ref{fig:mean-square_radius_fss}).
We make two key observations: 
(i) At early times, clusters spread diffusively with $z_s \,{=}\, 2$, followed by a crossover to an asymptotic critical behaviour with $z_s \,{\approx}\, 3$. (ii) Depending on the system size $L$, the mean-square radius $\langle R^2 \rangle$ saturates at some characteristic time scale $t_b(L)$.
This raises two central questions. What is the dynamic process that leads to this saturation and is it different from the process that drives the spreading of the cluster?
If the processes are different and both are associated with the critical dynamics of the DEP, then the data should obey the following \textit{generalised finite size scaling law}
\begin{equation}
    \langle R^2 \rangle (t,L) 
    =
    t^{2/z_s} 
    F (t/L^{z_b}) 
    \, , 
\label{eq:finite_size_scaling_R2}    
\end{equation}
with different dynamic exponents for \textit{cluster growth} ($z_s$) and the dynamics responsible for the \textit{saturation} of cluster growth ($z_b$). This is indeed what we find (Fig.~\ref{fig:mean-square_radius_fss}b): the scaling collapse works best for $z_b \, {=} \, 2.0(1)$ and $z_s \, {=} \, 3.0(1)$. 

\begin{figure}[!t]
\centering
\includegraphics[width=\columnwidth]{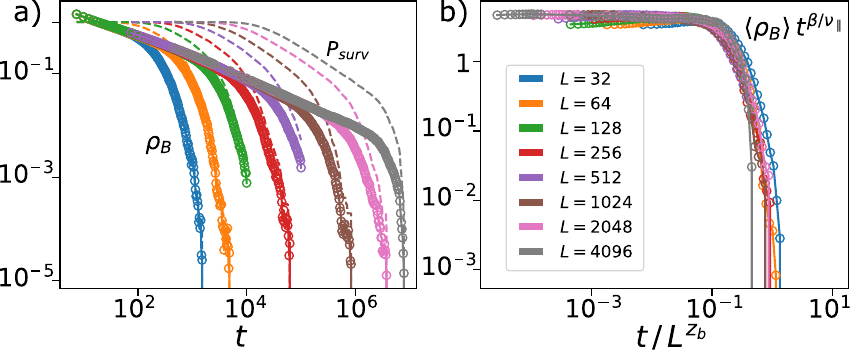}
\caption{
\textbf{Finite size scaling analysis for homogeneous initial state at criticality, $D_A > D_B$.}
(a) Simulation data for the density  $\langle \rho_B^{} \rangle$ (open circles) and the survival probability $P_\text{surv}$ (dashed lines) as a function of time $t$ at $\rho_c \,{=}\, 6.765$  and a set of system sizes $L$ as indicated in the graph. 
(b) A finite size scaling analysis for the density of $B$-particles using Eq.~\eqref{eq:finite_size_scaling_rhoB} produces a data collapse for $z_b \,{=}\, 2.0(1)$ and ${\beta}/{\nu_\parallel} \,{=}\, 0.46(3)$.
The ensemble size is $10 \, 000$.
}
\label{fig:hom_fss}
\end{figure}

We also performed a finite size scaling analysis of the simulation data obtained for systems initialised with a homogeneous distribution of $B$-particles. 
Figure~\ref{fig:hom_fss}a shows the time evolution of $\langle \rho_B \rangle (t)$  and $P_\text{surv}(t)$ at the critical density $\rho_c \,{=}\, 6.765$ for different system sizes.
The mean density exhibits a power law decay $\langle \rho_B^{} \rangle \,{\sim}\, t^{-\beta / \nu_\parallel}$ with the exponent $\beta / \nu_\parallel \,{=}\, 0.46(3)$ until finite size effects set in, which lead to an exponential decay into the absorbing state. 

The survival probability $P_\text{surv}$ exhibits two qualitatively different regimes. 
For small times,  $P_\text{surv} \,{\approx}\, 1$, indicating mean-field behaviour; 
note that the actual critical density $\rho_c$ is above the respective mean-field value $\rho_c^\text{mf} \,{=}\, 5$, so that there is a stable fixed point in this mean-field regime corresponding to a finite density of $B$-particles. 
The subsequent power law regime corresponds to critical behaviour with the same exponent as observed in the initial seed simulations, namely ${\beta'}/{\nu_\parallel} \,{=}\, 0.67(2)$. 
This is consistent with the intuition gained from individual simulation runs showing that an initially homogeneous spatial distribution of $B$-particles evolves into a set of well-separated $B$-clusters, suggesting that the survival of the active state is closely related  with the survival of individual $B$-clusters; compare Movie2 and Movie3.

The onset of finite size effects coincides for the survival probability and the $B$-particle density.
Similar as for the mean-square radius, we make the finite size scaling ansatz 
\begin{equation}
    \rho_B (L,t) 
    =
    t^{-\beta/\nu_\parallel} {\widetilde \rho_B} (t/L^{z_b})
    \, ,
\label{eq:finite_size_scaling_rhoB}   
\end{equation}
and find $z_b \,{=}\, 2.0(1)$; see Fig.~\ref{fig:hom_fss}b. 
This suggests that the characteristic processes that determine density fluctuations in systems initialised from a spatially homogeneous state are diffusive; we had already anticipated this by denoting this exponent as $z_b$. 
Taken together with the dynamic spreading data, this shows that there are --- in a sense yet to be specified --- two qualitatively distinct dynamic processes, one responsible for the spreading of clusters and the other for density fluctuations. 

\begin{figure}[tb]
\centering
\includegraphics[width=\columnwidth]{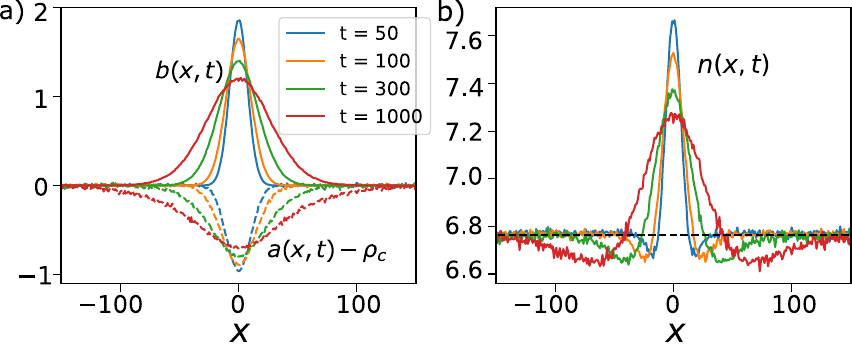}
\caption{
\textbf{Spreading dynamics at criticality, $D_A > D_B$.} Simulation data for systems (of size $L \,{=}\, 512$) initialised with a small cluster of $B$-particles at the origin, averaged over an ensemble of size of $100 \, 000$: Average densities of (a) $A$- and $B$-particles and of (b) the total density $n(x,t)$ for a set of times $t$ indicated in the graph.  
}
\label{fig:cluster_spreading}
\end{figure}

Consider a dynamic spreading simulation starting with a small $B$-cluster in a background of $A$-particles. 
Since $D_B \,{<}\, D_A$,  $B$-particles spread slower than $A$-particles.
Moreover, as sick individuals ($B$) infect healthy individuals ($A$) the `background' field $a(x,t)$ is reduced in the vicinity of the $B$-cluster; for an illustration see Fig.~\ref{fig:cluster_spreading}a and Movie2. 
Now, in stark contrast to the case of equal diffusion constants, the total density $n(x,t)$ shows a non-Gaussian profile: While in the center of the spreading cluster the density is above the critical density $\rho_c$, it then drops significantly below $\rho_c$ and approaches $\rho_c$ from below at large distances.
These `depletion zones' with $n(x,t) \,{<}\, \rho_c$ suppress the spreading of the $B$-cluster since they correspond to spatial regimes which are in the absorbing phase (Movie4). 
In this regime, the density of $B$-particles is driven exponentially fast to zero, creating a kind of `self-trapping' effect.
We hypothesize that this is the origin of the observed subdiffusive spreading with $\langle R^2 \rangle \,{\sim}\, t^{0.66(3)}$. 
%\bnote{It would be helpful if an effective model} for the cluster interface that makes the above heuristic arguments precise and allows to calculate the sub-diffusion exponent $2/z_s$ \bnote{could be found}. 

This leaves the question of why the saturation time for cluster growth scales as $t_b \,{\sim}\, L^2$. 
Since the $A$-particles spread faster than the $B$-particles ($D_A \,{>}\, D_B $) and the $B$-particles show self-trapping, this must be linked with the dynamics of the $A$-particles. 
Cluster growth depends on the influx of $A$-particles supplied by a diffusion process from the reservoir outside the cluster.
For an infinite system, this reservoir will not deplete and remain at the critical density $\rho_c$.
However, for a finite system, the reservoir will be depleted due to the continued influx of $A$-particles into the cluster on a time scale ${\sim}\, L^2/D_A$, so that the total density at the boundary of the system eventually falls below the critical density.
Once below the critical density the system will be driven exponentially fast towards the absorbing state as we observe in our simulations (Fig.~\ref{fig:mean-square_radius_fss}). In summary, the diffusive dynamics of the $A$-particles outside of the cluster drives saturation of cluster growth. 

Finally, the question remains how the dynamics reduces to one with a single time scale in the limit $D_B \,{\to} D_A$. 
Our simulations for $\langle R^2 \rangle$ show that for all $D_B \,{\neq}\, D_A$ there is the same crossover from diffusive ($z_s \,{=}\, 2$) to  subdiffusive ($z_s \,{=}\, 3$) spreading as found in  Fig.~\ref{fig:mean-square_radius_fss}a, with the crossover time $t_\times$ increasing as $D_B$ approaches $D_A$ (see SM~\cite{Note:SM}).
This suggests that $t_\times \,{\to}\, \infty$ as $D_B \,{\nearrow}\, D_A$ and that for unequal diffusion constants there is a crossover from a `diffusive' fixed point to a `strong-coupling' fixed point that has so far escaped renormalisation group analysis. 

In summary, our stochastic simulations show that in the DEP spreading of clusters containing sick individuals is subdiffusive with $z_s \,{\approx}\, 3$ while the density fluctuations of the background of healthy individuals remain diffusive. 
It is the extinction dynamics of these clusters - independent of the initial conditions - that dominates the survival of the active states and that determines the corresponding critical exponents.
These results strongly suggest that when diffusion constants are unequal with sick individuals spreading slower than healthy individuals, the dynamics are governed by a strong coupling fixed point. 
This puts the DEP process in the same class as other strongly coupling phenomena in non-equilibrium physics, such as surface growth~\cite{Kardar.Zhang.1986}.
We hope that our work will stimulate mathematical, possibly non-perturbative approaches that would help to decipher the observed anomalous dynamics.
\\

\begin{acknowledgments}
We would like to thank Fridtjof Brauns and Uwe T\"auber for stimulating and helpful discussions. 
We acknowledge financial support by the Deutsche Foschungsgemeinschaft through the Excellence Cluster ORIGINS under Germany's Excellence Strategy (EXC-2094-390783311). 
\end{acknowledgments}

%\bibliography{references_dep.bib}


\end{document}